\documentclass[letterpaper]{article} 
\usepackage{aaai23}  
\usepackage{times,amsmath}  
\usepackage{helvet}  
\usepackage{courier}  
\usepackage[hyphens]{url}  
\usepackage{graphicx} 
\usepackage{natbib}  
\usepackage{caption} 
\usepackage{algorithm}
\usepackage{algorithmic}

\usepackage{newfloat}
\usepackage{listings}
\DeclareCaptionStyle{ruled}{labelfont=normalfont,labelsep=colon,strut=off} 
\floatstyle{ruled}
\newfloat{listing}{tb}{lst}{}
\floatname{listing}{Listing}
\title{Unique in what sense? Heterogeneous relationships between multiple types of uniqueness and popularity in music}

\author{Yulin Yu\textsuperscript{1}\thanks{To whom correspondence may be addressed.}, Pui Yin Cheung\textsuperscript{2}, Yong-Yeol Ahn\textsuperscript{2}, Paramveer S. Dhillon\textsuperscript{1}}

\affiliations {
    \textsuperscript{\rm 1} School of Information, University of Michigan\\
    \textsuperscript{\rm 2} Indiana University Bloomington\\
    yulinyu@umich.edu, pcheung@iu.edu, yyahn@iu.edu, dhillonp@umich.edu
}

\usepackage{bibentry}

\begin{document}

\maketitle

\begin{abstract} 
 How does our society appreciate the uniqueness of cultural products? This fundamental puzzle has intrigued scholars in many fields, including psychology, sociology, anthropology, and marketing. It has been theorized that cultural products that balance familiarity and novelty are more likely to become popular. However, a cultural product's novelty is typically multifaceted. This paper uses songs as a case study to study the multiple facets of uniqueness and their relationship with success. We first unpack the multiple facets of a song's novelty or uniqueness and, next, measure its impact on a song's popularity. We employ a series of statistical models to study the relationship between a song's popularity and novelty associated with its lyrics, chord progressions, or audio properties. Our analyses performed on a dataset of over fifty thousand songs find a consistently negative association between all types of song novelty and popularity. Overall we found a song's lyrics uniqueness to have the most significant association with its popularity. However, audio uniqueness was the strongest predictor of a song's popularity, conditional on the song's genre. We further found the theme and repetitiveness of a song's lyrics to mediate the relationship between the song's popularity and novelty. Broadly, our results contradict the ``optimal distinctiveness theory'' (balance between novelty and familiarity) and call for an investigation into the multiple dimensions along which a cultural product's uniqueness could manifest.
\end{abstract}





\section{Introduction}

The uniqueness (or ``novelty'' or ``atypicality'') of a creative artifact is an important attribute that determines its artistic value. Products based on either a new combination of existing elements or groundbreaking principles~\cite{boden2004creative} can potentially lead to success. However, do consumers really prefer uniqueness in creative artifacts? Uniqueness in style, sound, or plot is often claimed to be a key determinant of an artist, song, or movie's artistic value and hence financial success. However, numerous pieces of music and other artistic works emphasize the importance of familiarity since a lack of novelty in artistic works is relatable to a wide range of audiences. This apparent tension between the association of novelty and familiarity with a cultural product's popularity is a source of significant debate among scholars. Unfortunately, strong empirical support for either side has been lacking~\cite{askin2017makes,berger2018atypical,jing2019sameness}. 

There are three competing theoretical explanations for the relationship between uniqueness and popularity of a cultural artifact. The first one highlights the role of imitation and similarity. In particular, it states that successful cultural products tend to imitate others by copying the ``winning formula'' and signaling conformity~\cite{aspers2013sociology,zuckerman2003robust}. 
A second theoretical explanation contradicts the importance of familiarity in determining the popularity and instead argues the critical role of the uniqueness of cultural products. It further suggests that uniqueness in cultural products stems from the creators' needs for expression and a desire to attract consumers by offering new experiences~\cite{jones2016misfits}. To reconcile these two conflicting arguments, many scholars have turned to a third explanation, called the optimal distinctiveness theory (also known as the balance theory). The optimal distinctiveness theory states that successful products have to be innovative enough to satisfy the consumers' need for a new experience and familiar enough to be comprehensible and accessible~\cite{berlyne1970novelty,goldberg2016fitting,lampel2000balancing,leonardelli2010optimal,zuckerman2016optimal}. 

The optimal distinctiveness theory has become an influential theory that has been supported by multiple empirical studies. For instance, scientific papers that strike a balance between novelty and typicality were found to be the most impactful~\cite{uzzi2013atypical}; when the novelty of movies was measured with keywords, it had an inverted U-shaped relationship with revenue~\cite{sreenivasan2013quantitative}. Along similar lines, participants found advertisements with moderate novelty to be the most persuasive~\cite{mohanty2016visual} and songs with moderately novel audio features were the most popular on Billboard charts~\cite{askin2017makes}. 
Yet, some studies contradict the optimal distinctiveness theory. For example, using a large sample of online fan-fiction ratings,~\citet{jing2019sameness} found the novelty of fan-fiction to be negatively associated with its popularity. They found a monotonically decreasing or a U-shaped relationship between novelty and popularity instead of the inverted U-shape suggested by the optimal distinctiveness theory.

A potential explanation for this conflict could be the definitional differences in describing a cultural product's novelty. Previous work has typically focused on a single dimension of uniqueness at a time, overlooking the multi-faceted complexity of cultural products. Cultural products such as music, visual art, and movies consist of multiple dimensions of expression. For instance, music includes elements such as lyrics, chord, sound (e.g., duration, pitch), rhythm, expression (e.g., dynamics, tempo, articulation), and melody~\cite{sarrazin2016music}. These dimensions are not necessarily correlated, and thus the ``uniqueness'' of a song can be wildly different depending on which dimension is used to define novelty. For example, a song's audio uniqueness does not necessarily imply its uniqueness in lyrics.

This raises several key questions:  Are these different uniqueness dimensions correlated? How is each novelty dimension associated with popularity, or which aspect of a song is the most important? Is there a universal relationship (e.g., the inverted U-shaped curve) between popularity and each dimension of novelty? For instance, one might expect optimal distinctiveness to be observed in a song's audio but not in its lyrics due to atypical lyrical themes being admired less by audiences.

This paper investigates the relationship between multiple dimensions of uniqueness and popularity in the domain of songs. We also study potential mechanisms through which a song's novelty could impact its popularity via a mediation analysis. More specifically, we ask the following research questions:

\noindent{\bf Research Question 1:} Does the relationship between uniqueness and popularity vary across the different uniqueness dimensions (i.e., audio, lyrics, or chord progressions) of a song? Which of these dimensions is the most important? 

\noindent{\bf Research Question 2:} What are the potential mechanisms linking a song's novelty and its popularity?

Our results have clear implications for social scientists and the music industry alike. Our findings can help the music industry understand the dynamics of music popularity and prepare effective marketing strategies. Content creators and songwriters can also benefit from our findings and tailor their songs to get broader attention.



\section{Empirical Setting and Data}

We assemble a novel dataset to measure multiple aspects of novelty and song popularity by combining resources from \url{Ultimate-guitar.com}, Gracenote API, and Spotify. 


\subsection{Data Description}


We use the lyrics and chords information for 123,837 songs collected by~\citet{kolchinsky2017minor} (who in turn collected it from \url{Ulitimate-Guitar.com} and GraceNote API). Next, we extracted the songs' audio features, e.g., danceability, energy, instrumentalness, loudness, speechiness, valence, tempo, song duration, and time signature, from the Spotify API. Finally, we gathered the popularity information of the songs and the artists, including each song's popularity score and the follower counts of the various artists. Our final dataset contains 51411 songs.\footnote{If a song had more than record on Spotify, we kept the one with the highest popularity. We also dropped songs with missing data.} Table~\ref{StatT} shows the summary statistics of our dataset stratified according to the region, period, and genre of songs in our dataset.
\begin{table*}[ht]
\centering
\tiny
\begin{tabular}{ll|ll|ll|ll|ll|lll}
\hline
\textbf{Region}   & \textbf{}                 & \textbf{Period}           & \textbf{}                 & \textbf{Genre}       & \textbf{}          
\ \textbf{}     & \textbf{Violence}       & \textbf{}          
& \textbf{Love}       & \textbf{}          
& \textbf{Dance}       & \textbf{}       \\
\hline
\hline
North America     & 33,760 & --1970 & 1,866  & Alternative \& Punk  & 19,256 & No & 43099 & No & 14143 & No & 49327\\
Western Europe    & 11,256 & 1971--1975                 & 1,294  & Classical/soundtrack & 124   &  Yes & 8312 & Yes & 37268 & Yes & 2084 \\
Australia/Oceania & 1,895  & 1976--1980                 & 1,195  & Electronica          & 566   &  &&&&&\\
Scandinavia       & 1,255  & 1981--1985                 & 1,074  & Jazz                 & 95   & &&&&& \\
Other/unknown     & 3,245  & 1986--1990                 & 1,683  & Other                & 4,231  &  &&&&&\\
                  &                           & 1991--1995                 & 3,016  & Pop                  & 8,139  & &&&&& \\
                  &                           & 1996--2000                 & 4,636  & Rock                 & 11,527 & &&&&& \\
                  &                           & 2001--2005                 & 9,190 & Traditional          & 5,995  &  &&&&&\\
                  &                           & 2006--2010                 & 16,409 & Urban                & 1,478  &  &&&&& \\
                  &                           & 2011--2015                 & 9,912 &                      &                           & &&&&&& \\
                  &                           & 2016--2020                 & 1,136  &                      &                           & &&&&& \\
\hline
\end{tabular}
 \caption{Summary statistics of categorical variables. {\it Note:} We perform our analyses at the level of songs.}
 \label{StatT}
\end{table*}

\subsection{Variable Construction}
We operationalize the three dimensions of a song's uniqueness\textemdash lyrics, chords, and audio\textemdash by comparing it to all the songs released in the same calendar year. We chose this operationalization since the uniqueness of a cultural product is time-dependent, and we need to control for time effects. Further, the novelty of cultural artifacts, e.g., songs, is often determined by the larger societal trends and ebbs-and-flows. Hence we examine the uniqueness of a song relative to its immediate temporal peers. That said, we performed robustness tests in which we varied the ``temporal comparison window,'' and our substantive results were consistent.

As one might expect, the relationship between song novelty and popularity should vary substantially across the different genres of the songs~\cite{askin2017makes}. So, we calculate the song uniqueness in two ways: (1) {\it Across-genre uniqueness:} comparing a song with all other songs regardless of genres and (2) {\it Genre-specific uniqueness:} comparing a song with all other songs in the same genre.

\paragraph{\bf a) Uniqueness Variables: } 

\begin{itemize}
    \item {\bf Lyrics Uniqueness:} We quantify the lyrics uniqueness by comparing the vector-space representations of songs from the same calendar year. Intuitively, a given song's lyrics would be less novel if they are highly similar to other songs' lyrics. We represent each song's lyrics as a vector whose entries are weighted by the term frequency-inverse document frequency (TF–IDF).\footnote{We also experimented with embeddings (Doc2vec) and topic modeling (LDA) (similar to \citet{aral-dhillon2023exactly} and \citet{jing2019sameness}) to quantify lyrics uniqueness. All three methods gave similar results, but we use TF-IDF owing to its simplicity.}

TF-IDF is a vector space modeling technique that uses a vector to represent each document. It downweights the commonly occurring terms in the document. In our case, a song is the equivalent of a document. Before generating the TF-IDF vectors of each song's lyrics, we preprocessed the raw text to remove non-English words, stopwords, and punctuation.

We first computed the TF-IDF representation of each song's lyrics. Next, we calculated the average TF-IDF of all the songs released in a given year. Finally, we obtained the lyrics uniqueness of each song by first computing the cosine similarity between a song's vector representation and the average vector representation of all the songs (excluding that song) from the same year and then subtracting it from 1. Equation~\ref{eq:reg112} defines the lyrics uniqueness $U^{Lyrics}_i$ of a given song $i$ that we use in our quantitative analyses later. $\text{Lyrics}_i$ denotes the TF-IDF vector representation of song $i$ and the set $S\backslash i$ is the set of all the songs released in the same year as the song $i$ excluding that song. When quantifying genre-specific lyrics uniqueness, we use the average TF-IDF of all the songs from a given genre released in a given year.




\begin{equation}
U_i^{\text{Lyrics}}= 1-\frac{\text{Lyrics}_i^{\top}\cdot \text{Lyrics}_{S\backslash i}}{\|\text{Lyrics}_i\|\|\text{Lyrics}_{S\backslash i}\|}
\label{eq:reg112}
\end{equation}



\item {\bf  Chord Uniqueness: }
When measuring chord uniqueness, we are interested in characterizing the frequency of individual chords in a given song instead of the full sequence of chords in that song. Once again, we represent each song's chords as a TF-IDF vector, where a song is the equivalent of a document and the specific chords, e.g., ``C Major'', ``E Minor'' etc., are the words in that document. As earlier, we first compute the cosine similarity of chord TF-IDF vector $\text{Chord}_i$ with the chord vectors of all songs released in the same year except that song $\text{Chord}_{S\backslash i}$. And then, we compute chord uniqueness by subtracting the chord similarity from one as shown in Equation~\ref{eq:reg111}.

\begin{equation}
U_i^{\text{Chord}}= 1-\frac{\text{Chord}_i^{\top}\cdot \text{Chord}_{S\backslash i}}{\|\text{Chord}_i\|\|\text{Chord}_{S\backslash i}\|}
\label{eq:reg111}
\end{equation}

\item {\bf Audio Uniqueness: }
As mentioned earlier, we collected the audio features of each song using the Spotify API. We followed the lead of~\citet{park2019global} in terms of what features to collect. We collected various audio features of each song, including danceability, energy, instrumentalness, loudness, speechiness, valence, tempo, song duration, and the time signature. The audio features were on different scales, so we rescaled them to lie between zero and one. Next, we concatenated them into a vector to compute the audio uniqueness in a similar fashion as the lyrics and chord uniqueness (see Equation~\ref{eq:reg113}).

\begin{equation}
U_i^{\text{Audio}}= 1-\frac{\text{Audio}_i^{\top}\cdot \text{Audio}_{S\backslash i}}{\|\text{Audio}_i\|\|\text{Audio}_{S\backslash i}\|}
\label{eq:reg113}
\end{equation}

\end{itemize}

\paragraph{\bf b) Popularity: }
Measuring the true popularity of a song is non-trivial. A song's popularity fluctuates over time and can differ wildly across populations. We use Spotify's track popularity metric for our analysis since streaming has become the dominant way of consuming music\footnote{Spotify is the largest music streaming platform globally with around 158 million subscribers as of March 2021.} We extracted this data in October 2020. We understand that this is a crude operationalization of popularity, but it is hard to find a better alternative, especially since we do not want to focus only on the most popular songs, e.g., Billboard top songs. In fact, a key goal of this paper is to study the long-tail of music consumption instead of just focusing on the most popular songs since they're not fully representative of the musical tastes and consumption of the masses. Hence, we employed the track popularity on Spotify as a proxy for the actual song popularity. According to Spotify's API documentation, ``The track's popularity is a value between 0 and 100, with 100 being the most popular. The popularity is calculated by an algorithm and is based, for the most part, on the total number of plays the track has had and how recent those plays are.''  It is worth noting that Spotify's popularity metric may lag the actual popularity by a few days since its value is not updated in real-time\footnote{https://developer.spotify.com/documentation/web-api/reference/}.

The three dimensions of song uniqueness (described earlier) and the Spotify song popularity are the four key variables of interest in our analyses. Their summary statistics are shown in Table~\ref{sumstat}.

\begin{table*}
\centering
\footnotesize
\def\sym#1{\ifmmode^{#1}\else\(^{#1}\)\fi}
\begin{tabular}{l*{5}{c}}

\hline
                &\multicolumn{1}{c}{$\mu$}&\multicolumn{1}{c}{$\sigma$}&\multicolumn{1}{c}{min}&\multicolumn{1}{c}{max}&\multicolumn{1}{c}{median}\\
\hline
Popularity (song)&    27.59 &        16.46          &    0              &   85 &    26 \\
Popularity (artist)&  1427995                &       4354019          &     0         &         75818586        &    232635         \\
Lyrics Uniqueness  &   0.83               &   0.07&        0          &    1&    0.84\\
Audio Uniqueness &    0.04 &        0.04          &           0       &  0.50 &    0.03 \\
Chord Uniqueness &  0.46                &       0.22           &     0         &         1         &     0.40         \\
Lyrics Uniqueness (genre-specific)  &   0.82               &   0.07&        0          &    0.99&    0.82\\
Audio Uniqueness (genre-specific)&    0.04 &        0.03          &           0       &  0.47 &    0.03 \\
Chord Uniqueness (genre-specific) &  0.45                &       0.22           &     0         &         1         &     0.40         \\
Sentiment &  0.34                &       0.73          &     -1         &         1        &    0.74 \\
Redundancy &  0.51                &       0.17          &     0.02         &         1        &    0.49         \\
\hline
\end{tabular}
\caption{Summary statistics of key variables}
\label{sumstat}
\end{table*}

Table~\ref{corr1} shows the Pearson correlation between the different types of song uniqueness. The relatively low correlation values suggest that the three dimensions of uniqueness capture non-redundant information of songs. The low correlation coefficients support the central premise of our study in measuring the multiple facets of song uniqueness independently. Table~\ref{corr2} presents the same correlation coefficients but for genre-specific uniqueness. Similarly, the correlations between the different types of uniqueness remain small, even if we control for the genres.  

\begin{table}[htbp]
\centering
\small
\begin{tabular}{l|l|l}
\hline
\hline
         & $\text{U}^{\text{Lyrics}}_i$                                           & $\text{U}^{\text{Chord}}_i$                    \\
\hline
$\text{U}^{\text{Chord}}_i$    & 0.06    &      -                    \\
\hline
$\text{U}^{\text{audio}}_i$ & 0.10 & 0.04   \\
\end{tabular}
\caption{Correlation between all-songs lyrics, chord, and audio uniqueness}
\label{corr1}
\end{table}

\begin{table}[htbp]
\centering
\small
\begin{tabular}{l|l|l}
\hline
\hline
         & $\text{U}^{\text{Lyrics}}_i$                                           & $\text{U}^{\text{Chord}}_i$                    \\
\hline
$\text{U}^{\text{Chord}}_i$    & 0.07    &      -                    \\
\hline
$\text{U}^{\text{audio}}_i$ & 0.12 & 0.04   \\
\end{tabular}
\caption{Correlation between genre-specific lyrics, chord, and audio uniqueness}
\label{corr2}
\end{table}

\paragraph{\bf c) Control variables:}
We include four sets of control variables to account for potential confounding in quantifying the impact of a song's lyrics, chord, or audio uniqueness on its popularity. 

\begin{itemize}
    \item {\bf Artist Popularity:}
    Artist popularity may affect the likelihood of writing unique songs. For instance, a popular artist may be more willing to take risks in experimenting with new features in songs than less popular artists. Also, it is easy to see that the artist's popularity would directly influence the song's popularity. Therefore, we need to control for artist popularity. We quantify an artist's popularity via the number of their followers on Spotify. The follower counts were extracted using the Spotify API\footnote{\url{https://developer.spotify.com/documentation/web-api/reference/#category-artists}}.
    
    \item {\bf Time Period:} Without a doubt, the release year of a song affects its popularity. The release year of a song also affects the features of the song. Hence, we include period dummies to control for such time effects. Each dummy variable corresponds to a five-year interval.

    \item {\bf Region:} There is also regional variation in the cultural production of music and music popularity. Songs produced in certain regions may be more well-received than others. Hence, we use region dummies to control for such ``location effects.'' We incorporated region dummies for North America, Western Europe, Australia/Oceania, Scandinavia, or other/unknown region.
    
    \item {\bf Genre:}
Finally, we control for the genre of the song. Once again, we operationalize this control variable via dummies. The songs in our dataset represented several genres, including alternative \& punk, classical, soundtrack, electronica, jazz, pop, rock, traditional, urban, or other/unknown genre. 
\end{itemize}

The descriptive statistics of all the control variables are shown in Table~\ref{sumstat}. 


\paragraph{\bf d) Song attribute variables:}
To probe potential mechanisms that may drive our results, we perform a mediation analysis. Based on the theory developed in previous works, we consider three potential mediators pertaining to the various song attributes.

\begin{itemize}
    \item {\bf Theme:}
The first potential mediator that we consider is the ``theme'' of the song. We calculate the theme of a song based on a word list developed by~\citet{berger2018atypical}. In addition, we also used some popular song themes listed in~\citet{christenson2019has}, e.g., ``Love and relationship,'' ``Dance,'' and ``Violence and Drugs.'' We coded the three theme variables as dummies where zero indicates the absence of the corresponding keyword and one indicates the presence of at least one related keyword. The keywords for each theme are listed below:

\begin{itemize}
    \item {\it Relationship and Love theme}: love, heart, burn, alive, feel, fire, hold, inside, light.
    \item {\it Dance theme}: bop, dab, funk, twerk, dance.
    \item {\it Violence and drugs theme}: dead gun, hate, kill, shit, slay, murder, shot, prison, drug.
\end{itemize}


  \item {\bf Sentiment:} We use the VADER algorithm to compute the sentiment of the song~\cite{hutto2014vader}. It outputs sentiment on a scale of [-1,1] with -1 indicating the most negative sentiment and +1 being the most positive sentiment. 

    \item {\bf Repetitiveness:} 
Finally, we consider the repetitiveness of a song's lyrics to mediate the impact of uniqueness on popularity since songs with repetitive lyrics are easy to remember and might become popular due to that. We coded repetitiveness as the proportion of unique words in a song's lyrics. For example, the repetitiveness of the following lyrics, ``Ice ice baby Ice ice baby All right stop,'' is 5/9. Thus, higher value of repetitiveness in our study refers to less duplicated words in lyrics.

\end{itemize}

\section{Empirical Models}
Our empirical analyses consist of several steps. First, we assess the relationship between a song's multi-faceted uniqueness and popularity using negative binomial regression model. After quantifying this relationship, we compare and contrast the importance of the specific dimensions of a song's uniqueness in determining its popularity. And, finally, we examine the potential mechanisms driving our results using mediation analysis.

\subsection{Relationship between Musical uniqueness and popularity: Negative Binomial model}

To answer our first research question, that is, quantifying the relationship between uniqueness and popularity, we estimate the model described in Equation~\ref{eq:nbreg11}. Since song popularity is a skewed and overdispersed count variable, we employ a negative binomial (NB) regression model to examine this relationship.

\begin{equation}
\text{Popularity}_i \sim \text{NB}(\text{U}^{\text{Chord}}_i+\text{U}^{\text{audio}}_i+\text{U}^{\text{Lyrics}}_i+\sum_k X_{ik}),
\label{eq:nbreg11}
\end{equation}
where, {$\text{Popularity}_i$} is the popularity of song $i$. Variables $\text{U}^{\text{Chord}}_i$, $\text{U}^{\text{Lyrics}}_i$ , and $\text{U}^{\text{Audio}}_i$ code the chord, lyrics, and audio uniqueness of the song $i$ respectively, as described earlier. Finally, $\text{X}_{ik}$ is the vector of control variables. 

\begin{table*}[htbp]\centering

\footnotesize
\def\sym#1{\ifmmode^{#1}\else\(^{#1}\)\fi}
\begin{tabular}{l*{5}{c}}

\hline\hline
                &\multicolumn{1}{c}{Model 6}&\multicolumn{1}{c}{Model 7}&\multicolumn{1}{c}{Model 8}&\multicolumn{1}{c}{Model 9}&\multicolumn{1}{c}{Model 10}\\
\hline
Audio Uniqueness ($\text{U}^{\text{audio}}_i$)&    -0.76\sym{***} &                  &                  &    -0.66\sym{***} &    -0.67\sym{***} \\
                &   (0.09)         &                  &                  &   (0.09)         &   (0.09)         \\
Lyrics Uniqueness ($\text{U}^{\text{Lyrics}}_i$) &                  &    -0.54\sym{***}&                  &    -0.51\sym{***}&    -0.51\sym{***}\\
                &                  &   (0.07)         &                  &   (0.07)         &   (0.08)         \\
Chord Uniqueness ($\text{U}^{\text{Chord}}_i$)&                  &                  &     0.00         &                  &     0.02         \\
                &                  &                  &   (0.02)         &                  &   (0.03)         \\
\hline
\emph{BIC}          &   401725                  &   401664                  &   401818                  &   401601                  &   401589                  \\
Observations    &    48766         &    48766         &    48766         &    48766         &    48766         \\
\hline\hline
\multicolumn{6}{l}{\scriptsize Standard errors in parentheses}\\
\multicolumn{6}{l}{\scriptsize Control variables include period, region, and genre. Coefficients of control variables are not shown here.}\\
\multicolumn{6}{l}{\scriptsize \sym{*} \(p<0.05\), \sym{**} \(p<0.01\), \sym{***} \(p<0.001\)}\\
\end{tabular}
\caption{Results of the Negative Binomial (NB) regression {for genre-specific uniqueness}. The number of observations in this analysis is smaller since certain periods and genres have insufficient sample sizes for calculating genre-specific uniqueness. Model 10 specifications provided in Equation~\ref{eq:nbreg11}.}
\label{regT-genre}
\end{table*}

\subsection{Mediation Analysis}
Next, we explore potential mechanisms explaining the dependence between lyrics uniqueness and popularity by performing mediation analyses. We consider three potential mediators here\textemdash theme, sentiment, and repetitiveness of the lyrics that can mediate the impact of uniqueness on song popularity. We believe that each of these mediators partially reflects the lyrical uniqueness of a song. For example, popular songs usually express themes such as love, dance, and violence~\cite{christenson2019has}. Similarly, a song's lyrics can be filled with positive sentiments such as happiness or negative emotion such as anger~\cite{dewall2011tuning}. Lastly, lyrical repetitiveness is a strong predictor of popularity~\cite{nunes2015power}. Hence, the three lyrical features represent nuanced information regarding a song and could potentially mediate the relationship between lyrics' uniqueness and popularity.

It turns out that there are even deeper motivations for choosing these mediators. First, repetitiveness has long been regarded by cognitive scientists as a key component of music similarity or uniqueness~\cite{volk2012towards}. Empirical evidence has shown that songs with high lyrics repetitiveness were more likely to succeed than those with low lyric repetitiveness~\cite{nunes2015power}. Thus, we postulated repetitiveness as one of the mediating mechanisms linking a song's lyrics uniqueness and popularity. Second, it turns out that the topical theme of a song is crucial in driving its popularity since popular songs center on familiar themes, e.g., ``love,'' ``violence,'' etc. Hence we hypothesized that the theme of a song could be a potential mediator between its lyrics uniqueness and popularity. Finally, a closer examination of the data reveals a prevalence of emotional valence in most songs. Hence we hypothesize that a song's sentiment could be a potential mediator. Thus, we tested the mediating effect of a song's lyrical theme (e.g., romantic, violence, or dance),  sentiment (e.g. positive or negative), and repetitiveness of lyrics.

\citet{mize2019general} proposed a general framework to perform mediation analysis. In particular, by exploiting Generalized Structural Equation Modeling's (GSEM) flexibility, multiple regression equations can be estimated simultaneously within the same GSEM model. A major advantage is that, under the GSEM framework, models with any link functions and distribution family can be estimated. Therefore, researchers can specify regression models with any type of dependent variable under GSEM. Moreover, regression predictions (e.g., coefficients, marginal effects) generated by different regression equations within the same GSEM model can be formally tested using Wald tests. Specifically, a Wald test statistic can be calculated by extracting the point estimates, the associated standard errors, and the covariance between point estimates. This test does not require the dependent variable to be continuous. As a result, this general framework can compare the difference in coefficients and any regression predictions across equations without restricting the type of variables involved.

We adopted this framework to perform mediation analysis in the context of our count outcome variable (popularity). Specifically, we examine five negative binomial regression equations simultaneously in the same GSEM model. The five models are shown in Equation 5. We add one group of mediators, i.e., theme, sentiment, repetitiveness, incrementally, and then estimate those different models simultaneously. For instance, the first model has no mediators; the following three models have theme, sentiment, and repetitiveness as mediator variables, respectively. And finally, the last model contains all the mediators at once. After estimating the entire GSEM model, we compared the regression coefficients of lyrics uniqueness using a series of Wald tests. Finally, we compute the reduction in coefficients of lyrics uniqueness and determine which group of mediators accounts for the relationship between lyrics uniqueness and popularity the most.

In the Equation 5, $\text{Popularity}_i$ is the popularity of the track $i$, $\text{U}^\text{Lyrics}_{i}$ is the lyrics uniqueness, and $\text{X}_{ik}$ is the vector of control variables. The $i$ subscript denotes an individual song. $\text{Theme}_{il}$ represents the vector of $l$ theme variables (i.e., violence, dance, and love), $\text{Sentiment}_{im}$ represents the vector of sentiment variables, and $\text{Repeat}_i$ denotes the repetitiveness of the lyrics.

\begin{eqnarray}
\nonumber
\text{Popularity}_i &\sim& \text{NB}(\text{U}^\text{Lyrics}_{i}+\sum_k\text{X}_{ik})\\
\nonumber
\text{Popularity}_i &\sim& \text{NB}(\text{U}^\text{Lyrics}_{i}+\sum_l\text{Theme}_{il}+ \sum_k\text{X}_{ik})\\
\nonumber
\text{Popularity}_i &\sim& \text{NB}(\text{U}^\text{Lyrics}_{i}+\sum_m\text{Sentiment}_{im}+ \sum_k\text{X}_{ik})\\
\nonumber
\text{Popularity}_i &\sim& \text{NB}(\text{U}^\text{Lyrics}_{i}+\text{Repeat}_{i}+ \sum_k\text{X}_{ik})\\
\nonumber
\text{Popularity}_i &\sim& \text{NB}(\text{U}^\text{Lyrics}_{i}+\sum_l\text{Theme}_{il}\\
\nonumber
&+&\sum_m\text{Sentiment}_{im}+\text{Repeat}_{i}+ \sum_k\text{X}_{ik})\\
\label{eq:gsem1}
\end{eqnarray}

\section{Results}

\subsection{Relationship between uniqueness and popularity}

Table~\ref{regT} reports the coefficients of the model specification described in Equation~\ref{eq:nbreg11}\footnote{We have tested the same models using bigrams and trigrams of lyrics and chord uniqueness. The substantive conclusions remain unchanged. The detailed results are presented in the robustness checks section.}. Models 1 to 3 include only one dimension of uniqueness (audio, lyrics, or chord) and the control variables. Model 4 includes only audio and lyrics uniqueness. Finally, Model 5 includes all the uniqueness dimensions together. Models 1-5 are designed to be increasingly complex to tease apart the impact of each type of song uniqueness on popularity by itself as well as together.

\begin{table*}[htbp]\centering
\small
\footnotesize
\def\sym#1{\ifmmode^{#1}\else\(^{#1}\)\fi}
\begin{tabular}{l*{5}{c}}
\hline\hline
                &\multicolumn{1}{c}{Model 1}&\multicolumn{1}{c}{Model 2}&\multicolumn{1}{c}{Model 3}&\multicolumn{1}{c}{Model 4}&\multicolumn{1}{c}{Model 5}\\
\hline
Audio Uniqueness ($\text{U}^{\text{audio}}_i$)&    -0.54\sym{***} &                  &                  &    -0.47\sym{***} &    -0.48\sym{***} \\
                &   (0.11)         &                  &                  &   (0.11)         &   (0.10)         \\
Lyrics Uniqueness ($\text{U}^{\text{Lyrics}}_i$) &                  &    -0.51\sym{***}&                  &    -0.49\sym{***}&    -0.50\sym{***}\\
                &                  &   (0.08)         &                  &   (0.08)         &   (0.09)         \\
Chord Uniqueness ($\text{U}^{\text{Chord}}_i$)&                  &                  &     0.01         &                  &     0.02         \\
                &                  &                  &   (0.03)         &                  &   (0.03)         \\
\hline
\emph{BIC}          &   424221         &   424100         &   424282         &   424073         &   424082         \\
Observations    &    51411         &    51411         &    51411         &    51411         &    51411         \\
\hline\hline
\multicolumn{6}{l}{\scriptsize Standard errors in parentheses}\\
\multicolumn{6}{l}{\scriptsize Control variables include period, region, and genre. Coefficients of control variables are not shown here.}\\
\multicolumn{6}{l}{\scriptsize \sym{*} \(p<0.05\), \sym{**} \(p<0.01\), \sym{***} \(p<0.001\)}\\
\end{tabular}
\caption{Results of the Negative Binomial (NB) regression for all-songs uniqueness by estimating Equation~\ref{eq:nbreg11}.}
\label{regT}
\end{table*}

We can make a few observations from the results. First, the coefficient of chord uniqueness is not statistically significant in both Models 3 and 5, which suggests that chord uniqueness may not be a useful dimension of uniqueness to predict popularity. Second, Models 1 and 2 have higher BIC (poorer model fit) than Models 4 and 5. This result suggests that while audio and lyrics uniqueness are both strong predictors of song popularity individually, they can jointly explain the popularity better than any one of them alone. Third, both audio and lyrics uniqueness have a relationship with song popularity that is statistically and economically significant. Fourth, Model 5 in Table~\ref{regT} shows that the coefficients of the three forms of uniqueness are similar to models with one dimension of uniqueness only (i.e., Models 1, 2, and 3). Therefore, this finding highlights the importance of taking various measurements of uniqueness into account since each form of uniqueness captures non-redundant and orthogonal information.


Interestingly, the coefficient of chord uniqueness is not significant (cf. Model 3 and Model 5). It has a negative association with popularity, unlike lyrics and audio uniqueness, which are positively related to popularity. Comparing the BIC from the five models (lower indicates a better fit), we see that Model 4 (no chord uniqueness predictor) has the best model fit. The results of Model 4 can be interpreted as follows\footnote{Since the uniqueness variables vary between 0 and 1, the standard interpretation based on one unit change in uniqueness would not be suitable. Thus, we interpret in 0.1 unit change.}: A 0.1 unit increase in audio uniqueness is associated with about 4.6\% decrease in popularity, controlling for all other variables (Model 4, p$<$0.05). Similarly, each 0.1 unit increase in lyrics uniqueness is associated with about 4.8\% decrease in popularity, controlling for all other variables (Model 4, p$<$0.05).

Broadly, our results suggest that lyrics uniqueness has a slightly stronger association with popularity than audio uniqueness, but the difference is not statistically significant (Wald test in Model 4, p$>$0.05). However, the chord uniqueness coefficient is significantly different (Wald test in Model 5, p$<$0.05) from the other two uniqueness measurements.

Next, we perform genre-specific analysis where we estimate the model described in Equation~\ref{eq:nbreg11} on the genre-weighted uniqueness variables. This analysis contains fewer data points since we could not compute genre-specific uniqueness for specific time-period and genre combinations due to insufficient sample size. In particular, our analysis did not include songs produced before 1970 and genres with fewer than 1000 observations, i.e., classical, electronica, and jazz songs. As a result, there were only 48,766 songs in this sub-analysis.

Table~\ref{regT-genre} reports the resulting model coefficients. The substantive conclusions remain largely the same as the models based on all songs. Once again, both audio and lyrics uniqueness are negatively associated with popularity (p$<$0.05), albeit with stronger associations. The coefficients of chord uniqueness in Models 8 and 10 are again not statistically significant. The model with audio and lyrics uniqueness (Model 9) and the model with all three uniqueness (Model 10) have the lowest BIC among the five models. Therefore, the result implies that genre-specific audio and lyrics uniqueness explain popularity well, and each of them provides an independent contribution to predicting popularity. On the other hand, chord uniqueness does not explain song popularity even after conditioning on the genre of the song. Even though the coefficient of audio uniqueness is larger in magnitude than that of lyrics uniqueness, their coefficients are not significantly different from each other (Wald test in Model 9, $p>0.05$). Thus, the predictive power of both genre-specific audio and lyrics uniqueness are similar.

We can interpret the Model 9 results as earlier\textemdash each 0.1 unit increase in audio uniqueness (genre-specific) is associated with about 6.4\% decrease in popularity, controlling for all other variables (Model 9, $p<0.05$). Similarly, each 0.1 unit increase in lyrics uniqueness is associated with approximately a 5.0\% decrease in popularity, controlling for all other variables (Model 9, $p<0.05$).

To summarize, we find varying degrees of negative association between popularity and multiple dimensions of uniqueness, highlighting the importance of examining various aspects of a song's uniqueness. Our analysis showed that audio and lyrics uniqueness consistently predict popularity in both with and without genre-weighted song comparisons.

\begin{figure}[h]
\centering
\includegraphics[width=.5\textwidth]{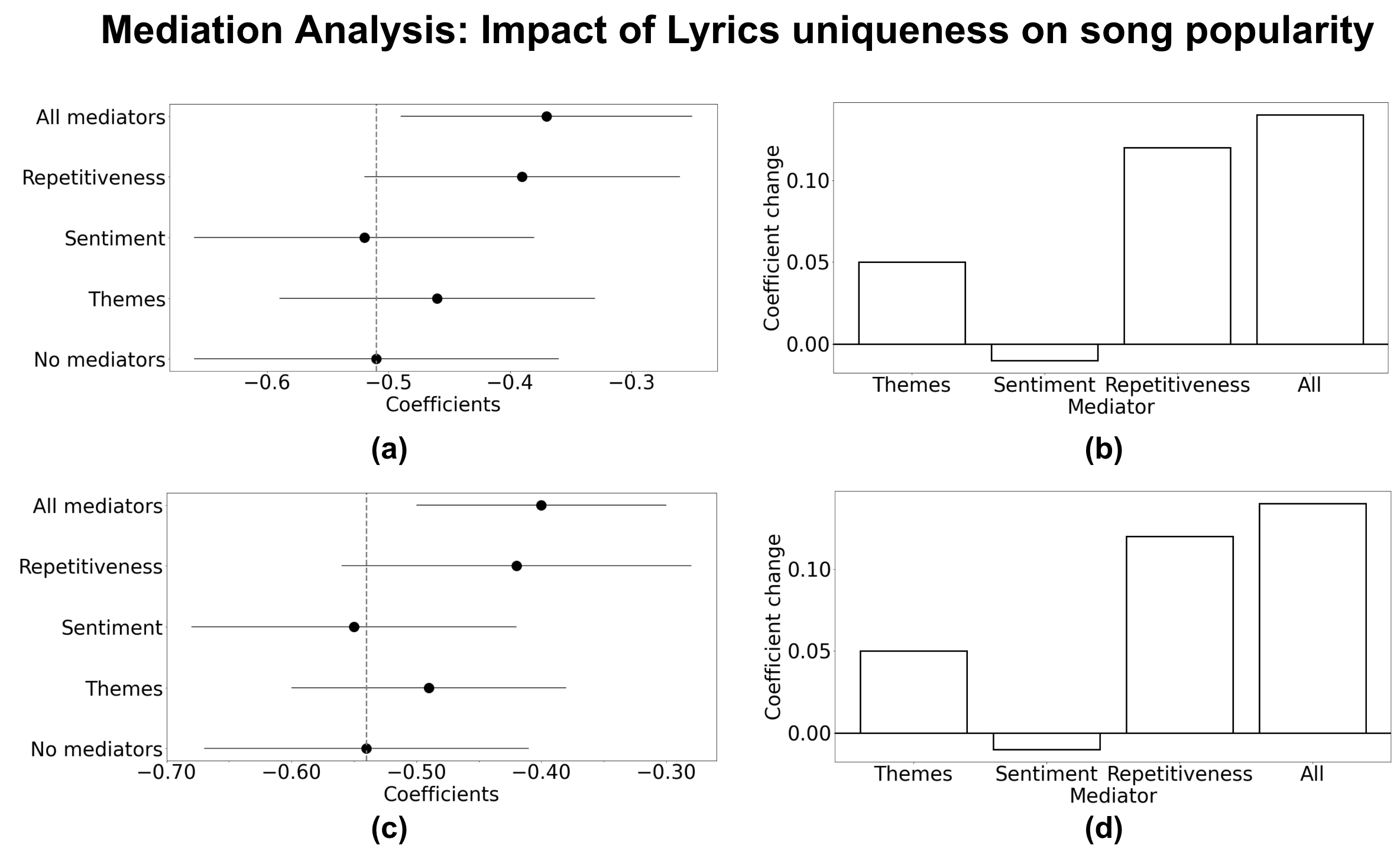}
\caption{(a) Coefficients of mediators: sentiment, theme, and repetitiveness. All mediators are significant. (b) The change in model coefficients of the lyrics uniqueness variable as the mediators are added incrementally.(c) Coefficients of mediators for genre-specific lyrics uniqueness: sentiment, theme, and repetitiveness. All mediators are significant. (d) The change in model coefficients of the lyrics uniqueness variable as the mediators are added incrementally.}
\label{fig:mediatorA}
\end{figure}

\begin{figure*}[htbp]
\centering
\includegraphics[width=.95\textwidth]{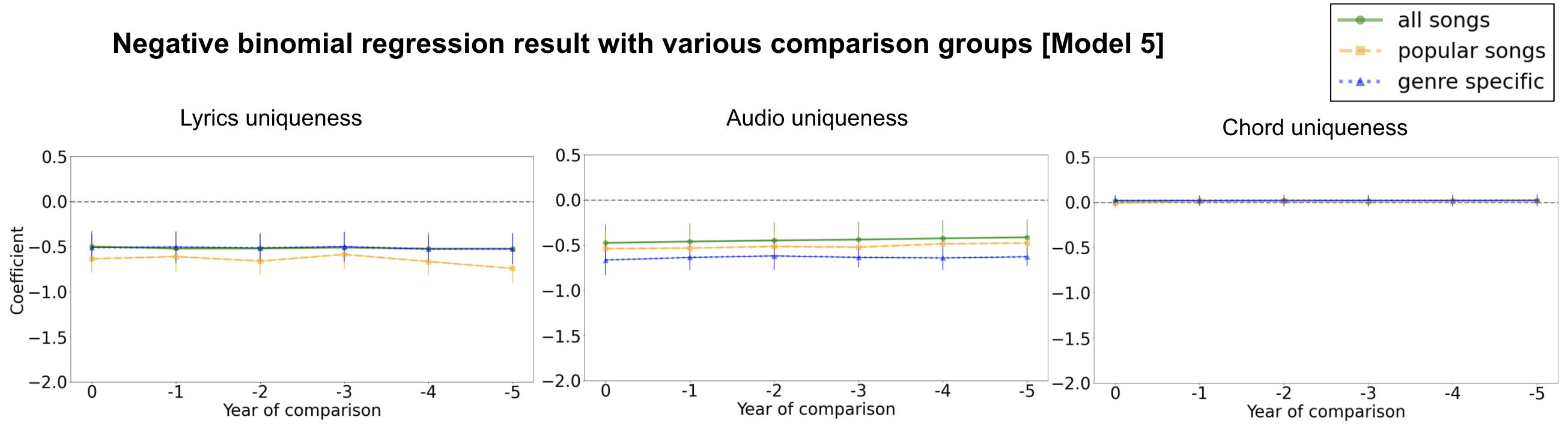}
\caption{Negative binomial regression model coefficients based on varying comparison groups (Model setting is the same as Model 5, Table \ref{regT}). {\it Note}: 1). The x-axis represents the comparison year for calculating uniqueness. For example, 0 indicates that uniqueness is calculated based on all songs released in the same year and -1 denotes uniqueness with respect to songs from the previous year. 2). ``Popular songs'' implies that uniqueness is calculated by comparing to popular songs (top 2.5\% popularity) released in the corresponding year. 3). The confidence intervals that are shown are 95\% intervals.}
\label{fig:regR}
\end{figure*}

\subsection{Mediation Analysis: Mediator of the impact of lyrics uniqueness on song popularity}

Next, we performed mediation analysis by estimating the GSEM model specified in Equation 5. In particular, we test the mediating impact of the theme, sentiment, and repetitiveness of the song's lyrics. The results are shown in Figure~\ref{fig:mediatorA}. The theme and repetitiveness of a song mediate the relationship between lyrics uniqueness and song popularity (p$<$0.05), whereas the sentiment expressed in the song's lyrics does not. The coefficient of lyrics uniqueness changes from -0.51 (without any mediators) to -0.46 (with theme as the mediator), to -0.39 (with repetitiveness as the mediator), and -0.37 (with all mediators). Even though the specification with sentiment as the mediator has a slightly higher point estimate (-0.52) than the equation without any mediators (-0.51), a Wald test does not indicate any significant difference between the two coefficients (p$<$0.05). Among the three mediators, repetitiveness is the strongest since the reduction in the point estimate of the coefficient is the largest. The difference in the absolute value of the coefficients of lyrics uniqueness between the equation with no mediators and the equation with all mediators is about 0.14 (p$<$0.05). This result suggests that two (theme, repetitiveness) out of the three mediators account for over a quarter of the association between lyrics' uniqueness and song popularity.

We observe a similar-sized mediating effect of the theme, sentiment, and repetitiveness of the song's lyrics when the lyrics uniqueness is measured just within the song genre. The results are visualized in Figure~\ref{fig:mediatorA}. In particular, the coefficient of genre-specific lyrics uniqueness changes from -0.54 (without any mediators) to -0.49 (with theme as the mediator), to -0.42 (with repetitiveness as the mediator), and -0.40 (with all the mediators). Similarly, the repetitiveness of a song's lyrics is the strongest mediator, and the mediating effect of the theme of the song is still significant. Overall, the three mediators explain about a quarter of the association between genre-specific lyrics uniqueness and song popularity.

\section{Robustness Checks}
We test the robustness of findings in several ways.

\subsection{ Validate lyrics uniqueness variable}
As a validation exercise, two undergraduate students manually labeled a randomly chosen set of songs' lyrics uniqueness. We randomly selected 10 songs each in the 1st and 4th quartile of lyrics uniqueness. Then, we paired all the unique combination of song from these groups and manually labeled which song is more unique. There are 100 uniqueness combinations in total. Our manual labels are consistent with labels generated by our algorithm with 85.7\% agreement, a Cohen's kappa of 0.6, and 74.5\% accuracy on average. The result suggests that our operationalization of uniqueness is reasonable.


\begin{figure}[htbp]
\centering
\includegraphics[width=.5\textwidth]{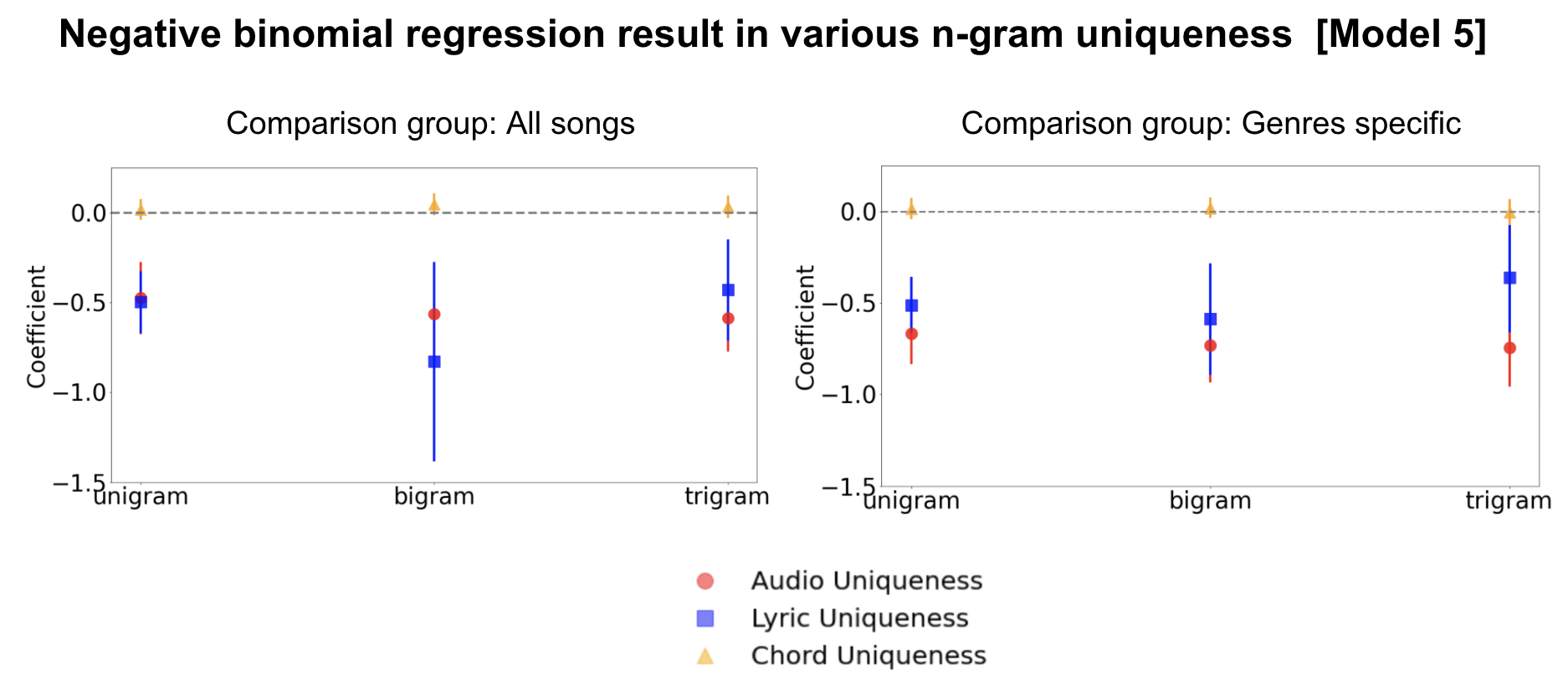}
\caption{Negative binomial regression model coefficients based on varying n-gram model regarding lyrics uniqueness and chord uniqueness (Model setting is the same as Model 5, Table \ref{regT}). {\it Note}: 1). The confidence intervals that are shown are 95\% intervals. 2)There are no n-gram variation in audio uniqueness}
\label{fig:regn-gram}
\end{figure}

\subsection{ Use n-gram to test word combination and chord progression uniqueness}

We calculated both lyrics and chord uniqueness with bigrams and trigrams. Since audio uniqueness is not estimated with n-gram, it remains unchanged throughout all n-gram settings. In each n-gram setting, both lyrics and chord uniqueness are included as the corresponding n-gram specification (i.e., unigram, bigram, trigram).  It is a common practice to use unigram, bigram,and trigram. n$>$3 is suitable to use when there is large training sample~\cite{jurafskyspeech}. In our dataset, since larger n would  mostly produce unique word combinations throughout the text, we didn't go beyond n=3. Figure~\ref{fig:regn-gram} presents the results by n-gram based on Model 5 in Table \ref{regT}. The point estimates of lyrics and chord uniqueness vary. However, they are substantively the same as unigram (negative and significant coefficients). As a result, we have decided to present and interpret the results based on unigram throughout the paper. Please note that the outcomes of genre-specific trigram analysis could be sensitive by the size of the sample within each genre. For instance, when "other genres" were excluded from the trigram analysis, resulting in a loss of about 4000 observations, the direction of point estimates remain the same as the main results. However, some regression coefficients in genre-specific uniqueness based on trigram became not statistically significant.

\subsection{Measuring quadratic relationship between music uniqueness and song popularity}
We tested other forms (potentially nonlinear) of the functional relationship between the different types of song uniqueness and popularity. In particular, we tested the presence of a quadratic relationship as posited by the optimal distinctiveness theory (inverted U-shaped relation)~\cite{askin2017makes}. However, the model with quadratic terms of uniqueness measurements does not fit better than the model with linear uniqueness terms only. Specifically, a model with quadratic terms of three dimensions of uniqueness (i.e., quadratic version of Model 5) has a BIC of 424084, a bit higher than Model 5's BIC. Due to the law of parsimony, we preferred the linear model (Model 5) over its quadratic version, despite both linear and quadratic versions produce substantively the same results. Moreover, the coefficients of quadratic terms were not statistically significant (except for audio uniqueness). Further, even with a statistically significant quadratic term, audio uniqueness had a curvilinearly decreasing relationship\footnote{After visualizing the marginal effects (not shown here) based on the empirical range of audio uniqueness, the curvilinear relationship is not distinguishable from a straight line. This further supports our conclusion that there is no curvilinear relationship.} with popularity rather than an inverted U-shaped relationship.

\subsection{Computing song uniqueness with different temporal comparison groups}

One might argue that a potential issue with our modeling approach is that highly popular songs may not look unique because other songs may immediately imitate them. In other words, the popular songs may have low uniqueness owing to their popularity, not the other way around. To rule out this possibility, we performed a robustness test by estimating Model 5 in Table~\ref{regT} with different comparison groups. In Table~\ref{regT}, we had calculated the song uniqueness by comparing it to all the songs released from the same year. So, we extended the year of comparison to n-5. In Figure \ref{fig:regR}, we calculated the uniqueness by comparing a song to all the songs released in n-1, n-2, n-3, n-4, or n-5 years (where n is the release year of that song). It is easy to verify that the coefficients at zero years of reference are identical to the Model 5 results in Table~\ref{regT}.

\begin{figure*}[htbp]
\centering
\includegraphics[width=0.95\textwidth]{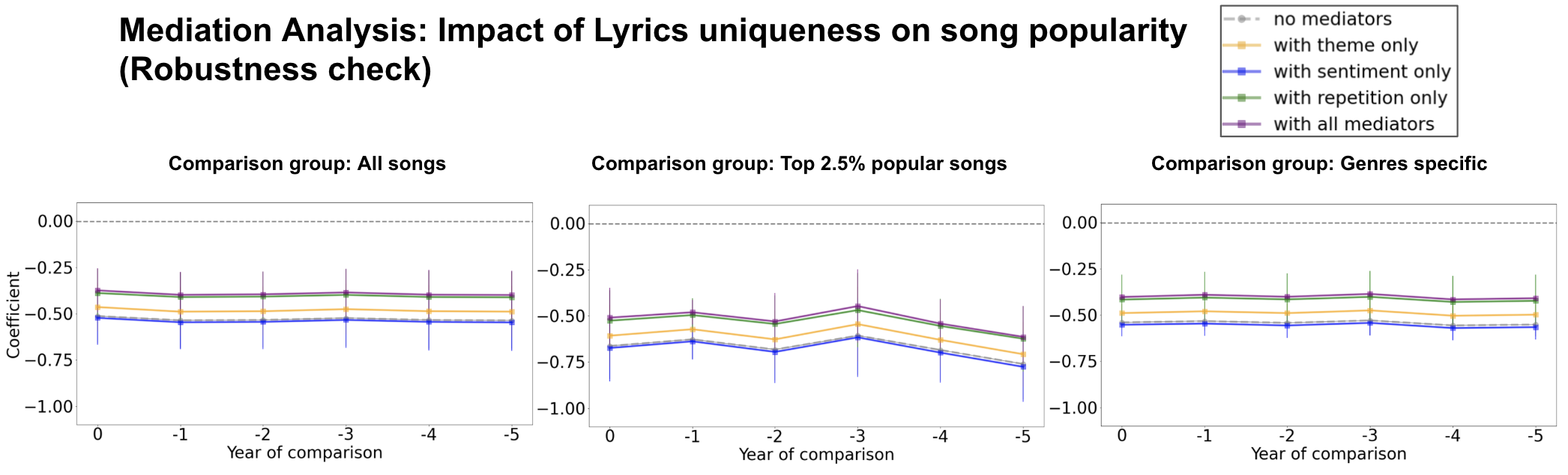}
\caption{Coefficients of lyrics uniqueness (Model setting is the same as the GSEM model in Figure \ref{fig:mediatorA}). The x-axis represents the comparison year for calculating lyrics uniqueness. For  example, zero  indicates that lyrics uniqueness is calculated based on all songs released in the same year, and -1 represents uniqueness computation with respect to songs from the previous year. The legend indicates the different regression model specifications (which correspond to Equation 5 above).}
\label{fig:mediatorR}
\end{figure*}

We performed the same robustness check for our mediation analysis as shown in Figure~\ref{fig:mediatorR}. In particular, we calculated the lyrics' uniqueness by comparing a song to songs released in different years. The results are qualitatively the same. In particular, adding theme, sentiment, and repetitiveness substantially reduces the coefficient of lyrics uniqueness across all comparison years. These results corroborate the robustness of our findings.

\subsection{Computing song uniqueness with reference to popular songs only}

Next, we tested the robustness of our findings by comparing a given song to only popular songs. We define ``popular'' songs as those with (Spotify) popularity scores higher than 60 in our dataset. We use this filter since songs with a popularity above 60 are in the top 2.5\% popular songs based on the $z$-score ($z$-score = 2). Our findings broadly remain the same. These results are shown in Figure~\ref{fig:regR} and Figure~\ref{fig:mediatorR}.

\subsection{Running separate analyses for each time period and genre}

One potential concern with our findings is that the results may not hold across periods or genres and might have a recency bias. This might be so since our popularity measurement favors the consumption of more recent songs than older songs since the users of streaming music apps are more likely to be from younger age groups. As a result, our substantive findings may not hold across all periods and genres. Hence, we perform a robustness test by running separate analyses using only one period or genre of songs at a time.

It turns out that the results by period and genre are broadly consistent with the main findings. In particular, audio and lyrics uniqueness are negatively associated with popularity (except a few genres and periods in which the coefficients are either not statistically significant or positive, such as the "other/unknown" genre). One major anomaly was that audio uniqueness had a significantly stronger relationship with popularity among traditional songs. Specifically, the coefficient of audio uniqueness is -2.46 in the model with traditional songs only. This result indicates that the relationship between uniqueness and popularity for songs in a particular genre (i.e., traditional) can be quite different from other genres. By and large, our analyses indicate that the relationship between uniqueness and popularity is substantively similar across periods and genres.

\subsection{Analyzing only recent songs}
\begin{figure}[htbp]
\centering
\includegraphics[width=.45\textwidth]{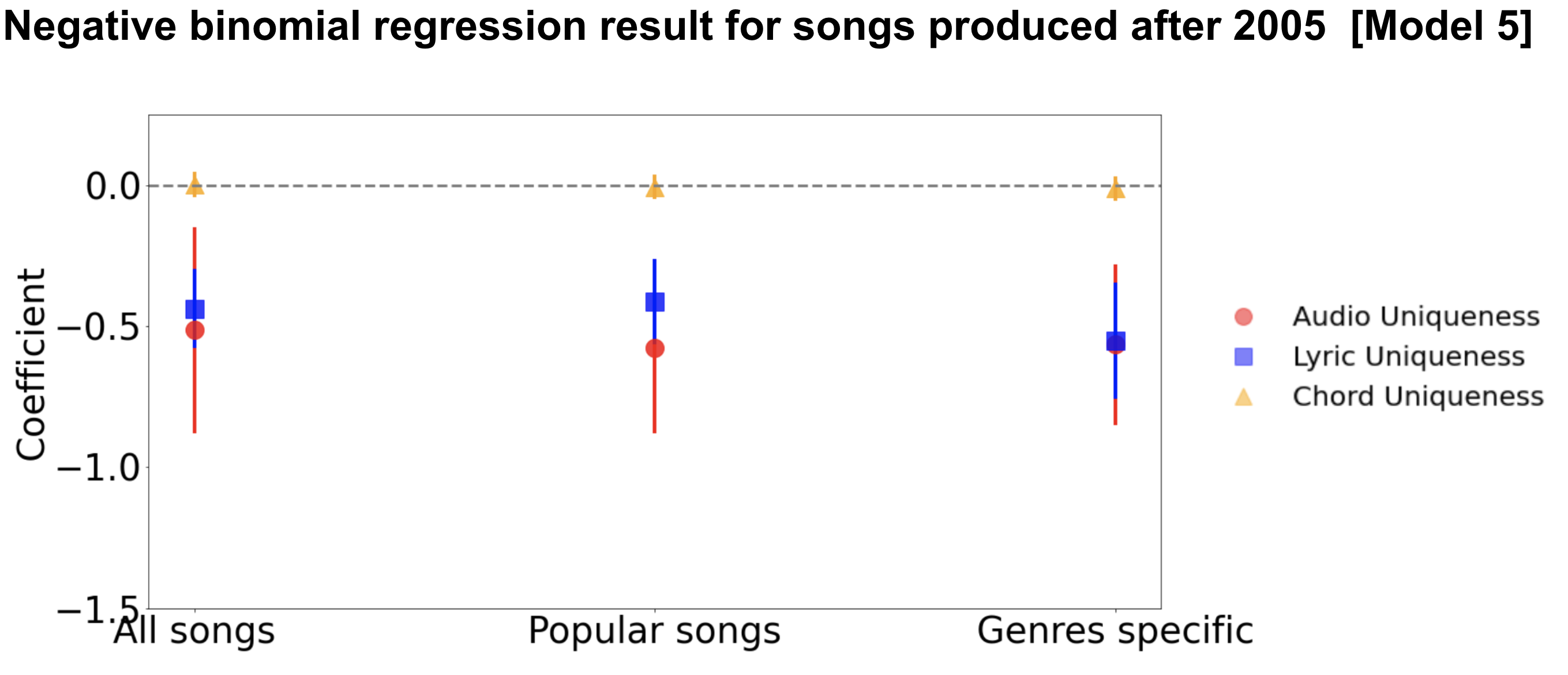}
\caption{Negative binomial regression model coefficients based on varying comparison groups on song created after 2005 (Model setting is the same as Model 5, Table \ref{regT}). {\it Note}: 1). The confidence intervals that are shown are 95\% intervals. }
\label{fig:regn-2005}
\end{figure}

Another way to address the heavier weights for more recent consumption of songs and the younger age groups of audiences is to perform analysis restricted to a more recent period, e.g., 2005 or later. It turns out that the coefficients based on the recent period are weaker than the main findings. Nevertheless, the coefficients of audio uniqueness and lyrics uniqueness remain significant and in the same direction in the restricted analysis. As a result, the substantive conclusion remains unchanged in the restricted analysis. These results are shown in Figure~\ref{fig:regn-2005}

\section{Discussion \& Limitations}

This paper investigated the relationship between a song's multidimensional uniqueness and its popularity. We hypothesized that various dimensions of music uniqueness, namely lyrics, chord, and audio, could determine its popularity and demonstrate that this is indeed the case by performing multiple statistical analyses. We then explored the nature of relationships between these dimensions of uniqueness and popularity. Finally, we examined the potential mechanism linking the lyrics uniqueness (one of the strongest predictors of song popularity) and popularity by considering mediators related to the song's theme, sentiment, and repetitiveness. 

In conclusion, we did not find evidence supporting the optimal distinctiveness theory. Instead, we discovered that uniqueness is negatively associated with popularity and that lyrics and audio uniquenesses have the strongest negative association with popularity. Further, we found that the repetitiveness of a song's lyrics is the strongest mediator of the relationship between uniqueness and popularity, rather than the theme or the sentiment expressed in the lyrics. The finding highlights the importance of taking multiple dimensions of songs into account.

There can be three possible explanations for the lack of evidence of the optimal distinctiveness theory. First, the current finding may be compatible with the optimal distinctiveness argument. In particular, empirical support for the optimal distinctiveness theory is based on highly competitive markets (e.g., the popular music industry). However, our research setting involves songs with diverse popularity. Hence, the market includes songs produced by indie artists and those supported by large and resource-rich corporations. Therefore, the dynamics of such a market would be quite different than the popular music industry. In an already highly competitive market (such as the popular music industry), striking a balance between typicality and uniqueness may help a song stick out and become one of the most popular songs. However, suppose we extend the scope of the market (i.e., a general music market where not all songs were created equal)\textemdash in that case, song popularity may be dependent on similarity to other songs, especially with popular songs. Future research should examine to what extent the dynamics of uniqueness and popularity measurement may explain the lack of optimal distinctiveness in our findings, compared to the previous studies that used sales or rank vary across market segments.

Second, the measurement of popularity in this study differs from previous studies. Specifically, our findings highlight that the similarity of lyrics and audio features are positively correlated with song popularity measured today. In contrast, previous studies usually employed Billboard chart ranking as the measurement of popularity, which is more contemporary to the release of the songs and focused only on the ``top of the charts.'' Our popularity measurement notably differs from Billboard chart ranking. In particular, there are many ways a song can be popular. For example, a song can be listened to by the same audiences multiple times, even after decades of its release. In other words, a popular song would be considered a ``classic'' by a group of people and be listened to repeatedly. Or a song can be a forgotten ``classic,'' i.e., it may be rediscovered by younger generations of audiences and undergo a revival. Billboard chart ranking cannot reflect these facets of popularity. In contrast, Spotify's popularity measurement can capture these multidimensional facets of popularity since it measures the cumulative audiences' listening behavior. Thus, this measurement discrepancy may explain the lack of optimal distinctiveness in our findings compared to previous empirical studies.

Third, it is worth noting that the anatomy of music consumption on streaming platforms such as Spotify is different than the consumption norms in the pre-streaming era. For instance, users often listen to playlists suggested by Spotify rather than constructing their own playlists. These Spotify playlists are generated algorithmically based on a user's tastes or based on similarity to the songs previously listened to by the user. Hence, users might not typically engage in the song selection process. Since Spotify playlists are generated based on some notion of similarity between songs, popular songs are likely to be present in many playlists. In other words, Spotify implicitly promotes music consumption based on similarity-based algorithmic recommendations. This could potentially alter the relationship between similarity (i.e., the opposite of uniqueness) and popularity. Hence, the idiosyncrasies of the music streaming platforms could potentially explain the difference between our main findings and the optimal distinctiveness theory.

Another explanation for the difference between our findings and the previous studies based on Billboard ranking is the duration of popularity. Songs that undergo a sudden spike in popularity and then fade out quickly can be captured by Billboard ranking. However, Spotify popularity is a cumulative measurement of music consumption which can fail to capture such fads.

The negative association between audio uniqueness and popularity indicates that people might seek familiarity in expression, including topics or values. Additionally, they might also be attracted to sound similarity. The result resonated with arguments in Schenker's theory~\cite{schenker2001free}. He argues that it is essential to have urlinie or musical motifs which repeat in different forms. He also mentions having respect for structure in music is important.

Lyrical theme and repetitiveness are strong mediators for the relationship between lyrics uniqueness and popularity. Among them, repetitiveness is the strongest mediator. Consistent with a previous study~\cite{nunes2015power}, we found that repetitiveness is negatively associated with popularity. Taken together, these mediators explain a significant portion of the relationship between lyrics uniqueness and popularity. Specifically, popular songs tend to be similar to other songs, and they can have a commonality in terms of expressed theme, expressed sentiment, or lyrics repetitiveness. We also found that repetitiveness mediates the association between uniqueness and popularity the most. This finding implies that songs with simpler lyrics may be more likely to become popular.

Our study has several limitations. First, the popularity metric we used was collected recently and based on Spotify's algorithm. This means that our popularity metric is determined by the \emph{current} users of Spotify and its algorithm. Therefore, the meaning of popularity for older songs is different from that of recent ones. Although we included period as a control variable and ran separate regression models by period, it is a potential limitation of our operationalization. Since Spotify's popularity metric reflects a cumulative music consumption weighted by recency of listening in the app \footnote{https://developer.spotify.com/documentation/web-api/reference/\#/operations/get-track}, the date of release of songs may affect the relationship between uniqueness and popularity. For instance, recent songs may not have been exposed to a large group of audiences compared to older songs. Thus, the relationships between uniqueness and popularity may differ between recently released and older songs. To address such concerns, we have performed robustness checks in Sections 5.3-5.6. The results indicate some differences in the relationships across release periods. However, the substantive conclusion remains the same. Second, our data contains more songs from 2006-2010 than 2016-2020. The difference in sample size is due to both copyright availability and the music industry's changing landscape. We performed a robustness check in Section 5.5 to address this concern about the difference in sample size. Fortunately, the substantive conclusion remains unchanged. Third, the sample of songs that we use in this study are sampled from the guitar tablature website and, therefore, may not represent all types of music. Although we introduced control variables such as genres and periods to alleviate this issue, it will be interesting to test our results in different datasets (e.g., hip-hop). Fourth, our study design only permits us to find associations, not causal relationships. Future work may employ experimental designs that manipulate the uniqueness to infer causal relationships. Finally, future work could also explore using more fine-grained linguistic models to extract even more nuanced linguistic features of music.

To summarize, our results suggest that (1) lyrics and audio uniqueness are most predictive of song popularity; (2) lyrics and audio uniqueness separately contribute significant information to a song's popularity; (3) these two dimensions of uniqueness are negatively associated with popularity linearly; (4) theme, sentiment, and repetitiveness explain a significant portion of the association between lyrics uniqueness and popularity; and (5) the song's lyrics' repetitiveness is the strongest mediator between the relationship of uniqueness and popularity, compared to the theme or the sentiment expressed in the lyrics. Collectively, these results highlight the importance of employing multiple uniqueness measurements to unpack how lyrics uniqueness can bring popularity. Our research shows that different aspects of multimedia's novelty (in our case, music) are differentially associated with popularity. This result also provides insight for future studies to understand the relationship between popularity and uniqueness in multimedia contexts besides music.

\section{Ethics Considerations}

The study was conducted on public data collected from various websites and Spotify API. We focus only on population-level music consumption data. Hence, there is no violation of individual-level privacy. However, the music listeners might not expect their data to be accessed by others via Spotify's API or scraped from the Internet.

\section{Acknowledgements}
We thank Daniel Romero, Misha Teplitskiy, Andrei Boutyline, 
Kerby Shedden, Elise Jing, and Jeff Lockhart for helpful discussions and feedback.

\newpage
\bibliography{cite}      

\end{document}